\documentclass[preprint,showkeys,aps,prb]{revtex4}
\usepackage{amsmath}
\usepackage[dvips]{graphicx}
\usepackage{setspace}

\begin{document}

\title{
Instantaneous Pairing of Lyapunov Exponents in Chaotic Hamiltonian Dynamics and the 2017 Ian Snook Prizes ;\\
Short Running-Head Title for CMST :\\ 2017 Snook Prizes : How Lyapunov Exponents Pair

}

\author{
William Graham Hoover and Carol Griswold Hoover       \\
Ruby Valley Research Institute  \\
HC 60 Box 601    \\
Ruby Valley, NV 89833                             \\
}

\date{\today}

\keywords{Chaos, Lyapunov Exponents, Algorithms}

\vspace{0.1cm}

\begin{abstract}
The time-averaged Lyapunov exponents, $\{ \ \lambda_i \ \}$, support a mechanistic description of
the chaos generated in and by nonlinear dynamical systems. The exponents are ordered from largest
to smallest with the largest one describing the exponential growth rate of the ( small ) distance
between two neighboring phase-space trajectories.  Two exponents, $\lambda_1 + \lambda_2$, describe
the rate for areas defined by three nearby trajectories.  $\lambda_1 + \lambda_2 + \lambda_3$ is
the rate for volumes defined by four nearby trajectories, and so on.  Lyapunov exponents for
Hamiltonian systems are symmetric. The time-reversibility of the motion equations links the growth
and decay rates together in pairs.  This pairing provides  a more detailed explanation than
Liouville's for the conservation of phase volume in Hamiltonian mechanics. Although correct for
long-time averages, the dependence of trajectories on their past is responsible for the observed 
lack of detailed pairing for the instantaneous ``local'' exponents, $\{ \ \lambda_i(t) \ \}$ . The
2017 Ian Snook Prizes will be awarded to the author(s) of an accessible and pedagogical discussion
of local Lyapunov instability in small systems. We desire that this discussion build on the two
nonlinear models described here, a double pendulum with Hooke's-Law links and a periodic chain of
Hooke's-Law particles tethered to their lattice sites.  The latter system is the $\phi^4$ model
popularized by Aoki and Kusnezov.  A four-particle version is small enough for comprehensive
numerical work and large enough to illustrate ideas of general validity.

\end{abstract}

\maketitle

\section{Introduction}

The elucidation of Hamiltonian chaos and Lyapunov instability by Poincar\'e and Lorenz is familiar
textbook material. Models which capture aspects of complexity, the Logistic and Baker Maps, the
Lorenz attractor and the Mandelbrot Set, combine visual appeal with mechanistic understanding in
the bare minimum of spatial dimensions, two for maps and three for flows. Mechanical models with
only three- or four-dimensional phase spaces are simple enough that the entire phase space can be
explored exhaustively.``Small Systems'' can augment our understanding of nature in terms of numerical
models by introducing more complexity. Just a few more degrees of freedom make an ergodic exhaustive
sampling impossible.  For the small systems we treat here we take on the more difficult task of
defining and analyzing the time-dependent convergence of ``typical'' trajectories. 

Chaos involves the exponential growth of perturbations.  Joseph Ford emphasized the consequence that
the number of digits required in the initial conditions is proportional to the time for which an
accurate solution is desired. Accordingly a ``typical'' nonexhaustive trajectory or history is the
best that we can do.  To go beyond the simplest models to those which elucidate macroscopic phenomena,
like phase transitions and the irreversibility described by the Second Law of Thermodynamics, we like
Terrell Hill's idea of small-system studies ( in the 1960s he wrote a prescient book,
{\it Thermodynamics of Small Systems}. ) In what follows we describe two small-system models which
are the foci of the Ian Snook Prize Problem for 2017.  These models are Hamiltonian, both with four
degrees of freedom so that their motions are described in eight-dimensional phase spaces.
\noindent

\begin{figure}
\includegraphics[width=4.0in,angle=-90.]{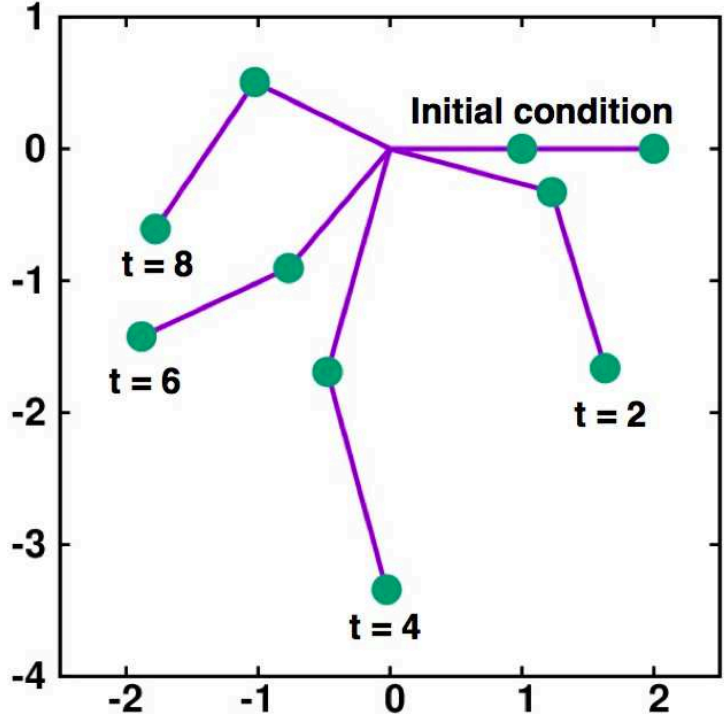}
\caption{
Snapshots of the two springy pendulum masses at times of 2, 4, 6, and 8.  Initially the pendulum is
horizontal as at the top right.  The initial configuration has vanishing energy. Only the outermost
particle, Particle 2, responds to the vertical gravitational field.\\
}
\end{figure}

\subsection{The Springy Pendulum and the Springy Double Pendulum}
The double pendulum with rigid links is an excellent model for the table-top demonstration of chaos.  Bill
saw one in action at an all-day Stanford lecture given by James Yorke.  An even simpler mathematical model
for chaos can be obtained with a single pendulum. For chaos the single pendulum needs a spring rather than
a rigid link.  The single springy pendulum moves in a four-dimensional phase space, just as does the
double pendulum with rigid links.  Along with Harald Posch\cite{b1,b2} we investigated mathematical models
for chaos based on chains of pendula, both rigid and springy.  We studied many-body instabilities by
characterizing the form of the detailed description of many-dimensional chaos, the Lyapunov spectrum.  We
considered two kinds of model Hamiltonians describing chains in a gravitational field :
[ 1 ] chains composed of particles
with equal masses, as in a physical length of chain; [ 2 ] chains in which only the bottom mass was affected
by gravity, as in a light chain supporting a heavy weight. Figure 1 shows five snapshots, equally spaced in
time, from a chaotic double-pendulum trajectory.  Initially the motionless chain was placed in the
horizontal configuration appearing at the top right of Figure 1.  If gravity affects only the lower of the
two masses ( as in the type-2 models supporting a heavy weight ) the  corresponding Hamiltonian is
$$
{\cal H} = [ \ p_1^2 + p_2^2 \ ]/2 + (\kappa/2)[ \ (r_1-1)^2 + (r_{12}-1)^2 \ ] + y_2 \ .
$$
where $r_1$ and $r_{12}$ are the lengths of the upper and lower springs. To enhance the coupling between
the springs and gravity we choose the force constant $\kappa = 4$ here.\\

\subsection{The Spectrum of Time-Averaged Lyapunov Exponents, $\{ \ \lambda \ \}$ }

The Lyapunov exponents making up the spectrum are conventionally numbered in the descending order of their
long-time-averaged values. We begin with the largest, $\lambda_1$ . $\lambda_1$ describes
the long-time-averaged rate at which the distance between the trajectories of two nearby phase-space
points increases. That rate,
$\lambda_1 \equiv \langle \ \lambda_1(t) \ \rangle \equiv \langle \ (d\ln\delta/dt) \ \rangle$ ,
is necessarily positive in a chaotic system.  A more detailed description of rates of change of lengths and areas, and
volumes, and hypervolumes of dimensionality up to that of the phase space itself, leads to definitions
of additional Lyapunov exponents.  The next exponent, $\lambda_2$ , is needed to describe the rate at which
a typical phase-space  area, defined by three nearby points, increases ( or decreases ) with increasing time,
$\lambda_1 + \lambda_2 \equiv \langle \ (d\ln A/dt) \ \rangle =
\langle \ \lambda_1(t) + \lambda_2(t) \ \rangle \ $.
Again an average over a sufficiently long time for convergence is required.  Likewise the time-averaged
rate of change of a three-dimensional phase volume defined by four neighboring trajectories is $\lambda_1
+ \lambda_2 + \lambda_3$ . This sequence of rates and exponents continues for the rest of the spectrum.
There are $D$ exponents for a $D$-dimensional phase-space description.

\subsection{Local and Global Lyapunov-Exponent ``Pairing'' for Hamiltonian Systems}

The time-reversibility of Hamiltonian mechanics implies that all the rates of change change sign if the
direction of time is reversed.  This suggests, for instance, that all the exponents, $\{ \ \lambda \ \}$ and
$\{ \ \lambda(t) \ \}$ , are ``paired'', with the rates forward in time opposite to those backward in
time.  This turns out to be ``true'' for the long-time-averaged exponents but could be ``false'' for the local
exponents.
Local exponents depend upon the recent past history of neighboring trajectories.  The global exponents,
which describe the growth and decay of the principal axes of comoving hyperellipsoids in phase space are paired,
though the time required to show this through numerical simulation can be long.  This exponent pairing is
the focus of the 2017 Snook Prize, as we detail in what follows.  There is a vast literature describing and
documenting the numerical evaluation and properties of Lyapunov spectra.  The theoretical treatments are
sometimes abstruse and lacking in numerical verification.  This year's Prize Problem seeks to help remedy this
situation.  The numerical foundation for the study of Lyapunov
exponents is an algorithm developed by Shimada and
Nagashima in Sapporo\cite{b3} and Benettin in Italy, along with his colleagues Galgani, Giorgilli,
and Strelcyn\cite{b4}, beginning in the late 1970s.  Google indicates hundreds of thousands of internet
hits for ``Lyapunov Spectrum''.  We mention only a few other references\cite{b5,b6,b7,b8} here. The
internet makes these and most of the rest readily available.\\

\subsection{The $\phi^4$ Model for Chaos and Heat Conduction in Solids}
\noindent

\begin{figure}
\includegraphics[width=4.5in,angle=-90.]{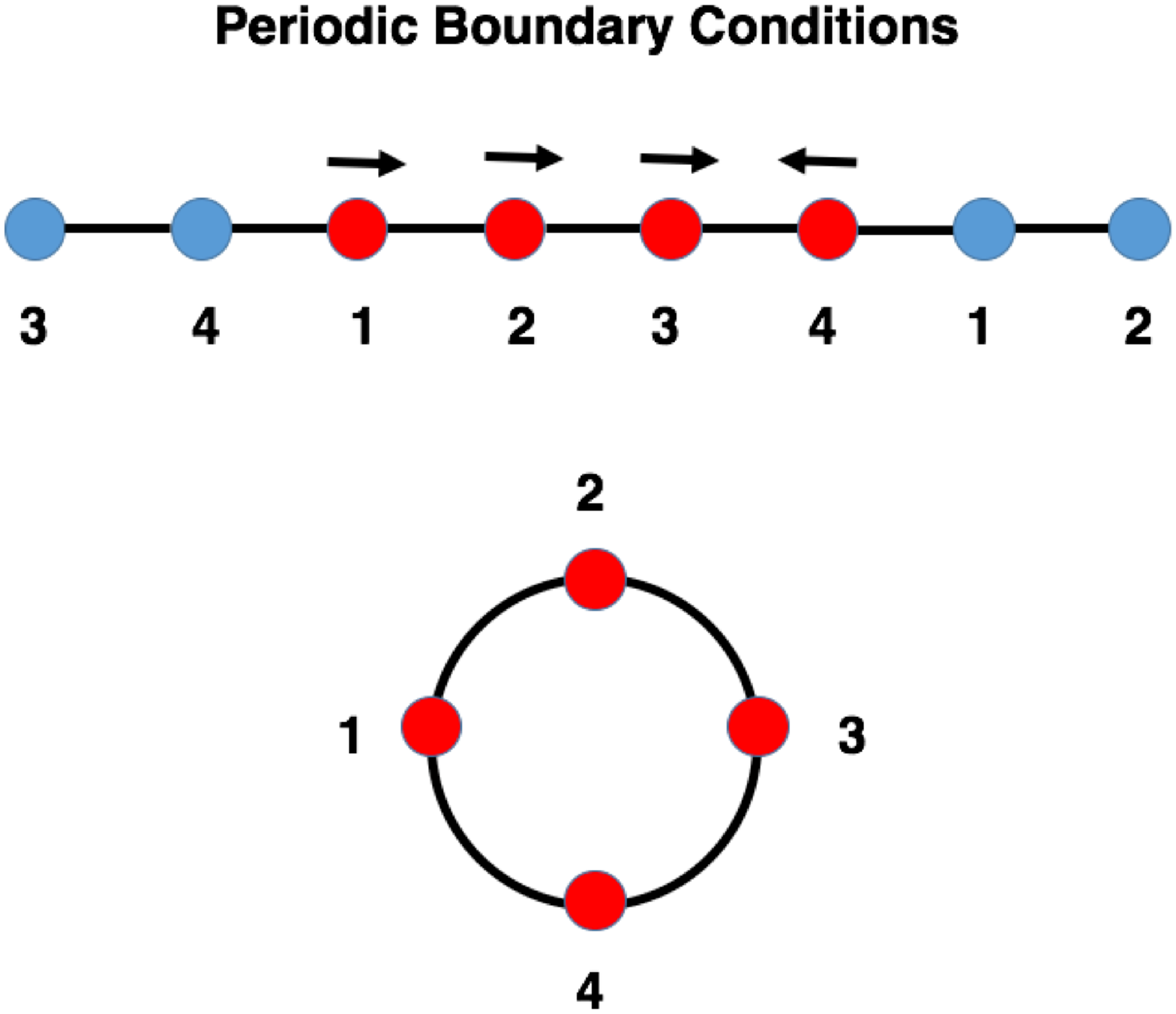}
\caption{
The four-body $\phi^4$ chain can be pictured as an infinitely-long chain with a four-particle
unit cell ( shown in red ), or as an arrangement of masses on a circle with equally-spaced lattice sites to which the
particles are tethered.  It is convenient ( but not foolproof ) to choose an initial condition with velocities
$
\{ \ +v,+v,+v,-v \ \} \ {\rm or} \ \{ \ +2v, 0,0,0 \ \}\longrightarrow (E/N) = (v^2/2) \ .
$
The initial conditions $\{ \ +v,+v,-v,-v \ \} \ {\rm and} \ \{ \ +v,-v,+v,-v \ \}$ are ``normal modes'',
discussed at length in References 10 and 11.
}
\end{figure}

Aoki and Kusnezov popularized the $\phi^4$ model as a prototypical atomistic lattice-based model leading to
Fourier heat conduction\cite{b9,b10,b11}.  In addition to a nearest-neighbor Hooke's-Law potential the
model incorporates quartic tethers binding each particle to its own lattice site.  

Here we denote the
displacements of the particles from their sites as $\{ \ q_i \ \}$ . In our one-dimensional
case the spacing between the lattice sites does not appear in the Hamiltonian or in the equations of
motion.  In numerical work it is convenient to choose the spacing equal to zero while setting the
particle masses, force constants for the pairs, and those for the tethers all equal to unity.  For a
four-particle problem in an eight-dimensional phase space the three-part Hamiltonian is :
$$                                                                                                                 
{\cal H} = \sum_{i=1}^4 [ \ (p_i^2/2) + (q_i^4/4) \ ] + \sum_4^{springs} (q_{i,j}^2/2) \ .                         
$$
The periodic boundary condition includes the spring linking particles 1 and 4 :
$$                                                                                                                 
\ddot q_1 = -q_1^3 + q_2 + q_4 - 2q_1 \ ; \ \ddot q_4 = -q_4^3 + q_1 + q_3 - 2q_4 \ .
$$
See Figure 2 for two ways of visualizing the periodic boundary conditions of the $\phi^4$ chain.\\
The energy range over which chaos is observed in the $\phi^4$ model includes about nine orders of
magnitude\cite{b10,b11}.  The chaotic range for a four-body chain includes the two cases we discuss in
the present work, $\{ \ E = 8, \ 288 \ ; \ (E/N) = 2,\ 72 \ \} $ . With both the springy pendulum and
the $\phi^4$ models in mind we turn next to a description of their chaotic properties.

\section{The Chaotic Dynamics of the Springy Double Pendulum}
 
Like most smoothly-differentiable Hamiltonian systems the double springy pendulum has infinitely
many periodic  or quasiperiodic phase-space solutions surrounded by a chaotic sea. Dynamics in the
sea is exponentially sensitive to perturbations.  The dynamics occurs in an eight-dimensional phase
space. Perturbations oriented along the trajectory or perpendicular to the energy surface, where there
is no longtime growth at all, give two zeroes, so that the maximum number of nonzero Lyapunov
exponents is six.

Each positive exponent is necessarily paired with its negative twin, with the two changing roles if
the direction of time is reversed. It is often stated that this time-reversible pairing links not
only the time-averaged rates of the dynamics, but also the ``local'' or ``instantaneous'' rates\cite{b2}. 
Because chaotic pendulum problems give different local exponents if Cartesian and polar coordinates
are used one might think that pairing could be hindered by using a mixture of these coordinates.
To check on this idea we considered a mixed-coordinate Hamiltonian for the model of Figure 1 with
polar coordinates for the ``inside'' Particle 1 :
$$
{\cal H} = (1/2)[ \ p_r^2 + (p_\theta/r)^2 + p_x^2 + p_y^2 \ ] + y_2
+ (\kappa/2)[ \ (r-1)^2 + (r_{12}-1)^2 \ ] \ ;
$$                                                                                                                 
$$
r_{12} = \sqrt{ x_2^2 + y_2^2 + r_1^2 - 2r_1x_2\sin(\theta_1) + 2r_1y_2\cos(\theta_1)} \ ; \ \kappa = 4 \ .
$$
Formulating and solving the motion equations in mixed Cartesian and polar coordinates is an
intricate error-prone task.  It is useful first to solve the problem in Cartesian coordinates.
That solution then provides a check for the more complicated mixed-coordinate case. Energy
conservation is a nearly-infallible check of the programming.  We computed spectra of
Lyapunov exponents averaged over one billion fourth-order and one billion  fifth-order Runge-Kutta
timesteps, $dt = 0.001 $ . This ensures that the numerical truncation errors of order $(dt^5/120)$ or
$(dt^6/720)$ are of the same order as the double-precision roundoff error. We chose the initial
condition of Figure 1 with both masses motionless at the support level, $\{\ x_1,y_1,x_2,y_2\ \} =
\{ \ 1,0,2,0 \ \}$ , so that the initial potential, kinetic, and total energies all vanished.
Only the outer Cartesian mass interacts with the gravitational field.

The simplest numerical method for obtaining Lyapunov spectra\cite{b3,b4} is first to generate a
$D$-dimensional ``reference trajectory'' in the $D$-dimensional phase space.  Then a set of $D$
similar ``offset'' trajectories, an infinitesimal distance away, $\delta$ , are generated in
the same space with numerical offset vectors of length $\delta = 0.00001$ or 0.000001.  While advancing the
resulting $D(D+1)$ $D$-dimensional differential equations the local Lyapunov exponents are
obtained by ``Gram-Schmidt'' orthonormalization. This process rescales the vectors to their original
length and rotates all but the first of them in order to maintain their orthonormal arrangement.
The rescaling operation portion of the Gram-Schmidt process gives local values for the $D$ Lyapunov
exponents :
$$
\lambda_i(t) \equiv (-1/dt)\ln(\delta_i^{after}/\delta_i^{before}) \ ; \ \lambda_i \equiv 
\langle \ \lambda_i(t) \ \rangle \ . 
$$
For the type-2 double pendulum of Figure 1 the time-averaged Lyapunov spectrum is :
$$
\{ \ \lambda \ \} = \{ \ +0.143, +0.076. +0.034, 0.000, 0.000, -0.034, -0.076, -0.143 \ \} \ .
$$
The rms fluctuations in these rates are typically orders of magnitude larger than the rates themselves. 
The uncertainty in the exponents as well as the differences between exponents using fourth-order
or fifth-order Runge-Kutta integrators with $dt = 0.001$ are both of order $\pm 0.001$ . Our numerical work
shows that the pairing of the exponents is maintained if one of the pendula is described by polar
coordinates with the other pendulum Cartesian.  The local exponents are different but still paired.\\

\noindent

\begin{figure}
\includegraphics[width=5.0in,angle=+90.]{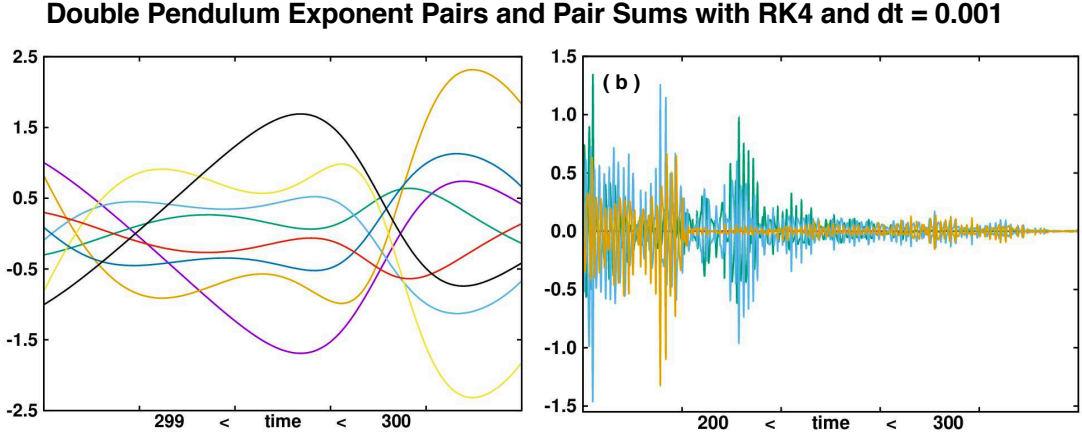}
\caption{
Typical pairing (a) of the eight Lyapunov exponents for the double-pendulum problem using Cartesian
coordinates stored in the order $\{ \ (x,y,p_x,p_y)_1,(x,y,p_x,p_y)_2 \ \}$ , with the $i$th offset
vector initially parallel to the $i$th member of this set of variables. Typical equilibrating pair sums
$\{ \ \lambda_i(t) + \lambda_{9-i}(t) \ \}$ are also shown in (b).  Fourth-order double-precision Runge-Kutta
integration was used with a timestep $dt= 0.001$ and offset vectors of length $\delta = 0.000001$ .
The initial condition is that shown in Figure 1.\\
}
\end{figure}

\section{Convergence and Ordering of Local Lyapunov Exponents}

The algorithm for generating the Lyapunov exponents\cite{b3,b4} requires the ordering of $D$ offset
vectors in the vicinity of a reference trajectory.  The first vector follows exactly the same motion
equations with the proviso that its length is constant.  The second vector, also of constant length,
is additionally required to remain orthogonal to the first so that the combination of the two gives
the rate of expansion or contraction of two-dimensional areas in the vicinity of the reference
trajectory. In general the $n$th offset vector satisfies $n$ constraints in all, keeping its own
length constant while also maintaining its orthogonality to the preceding $n-1$ vectors.

Although the local rates $\{ \ \lambda(t) \ \}$ associated with the vectors are necessarily ordered
when time-averaged  over a sufficiently long time to give the $\{ \ \lambda \ \}$ , this
ordering is regularly violated, locally, as Figures 3 and 4 show. Offhand one would
expect that increasing the Lyapunov exponents or decreasing the accuracy of the simulation would
lead to more rapid convergence of the ordering of the vectors.  For this reason we consider a
model which is as simple as possible, with a relatively large chaotic range, and is easy to simulate.
This $\phi^4$ model, named for its quartic tethering potential, has proved particularly useful in
the simulation of heat flow.  We consider the equilibrium version of the model here, an isolated system.

\section{The Dynamics of One-Dimensional Periodic $\phi^4$ Models}

\noindent

\begin{figure}
\includegraphics[width=5.5in,angle=-90.]{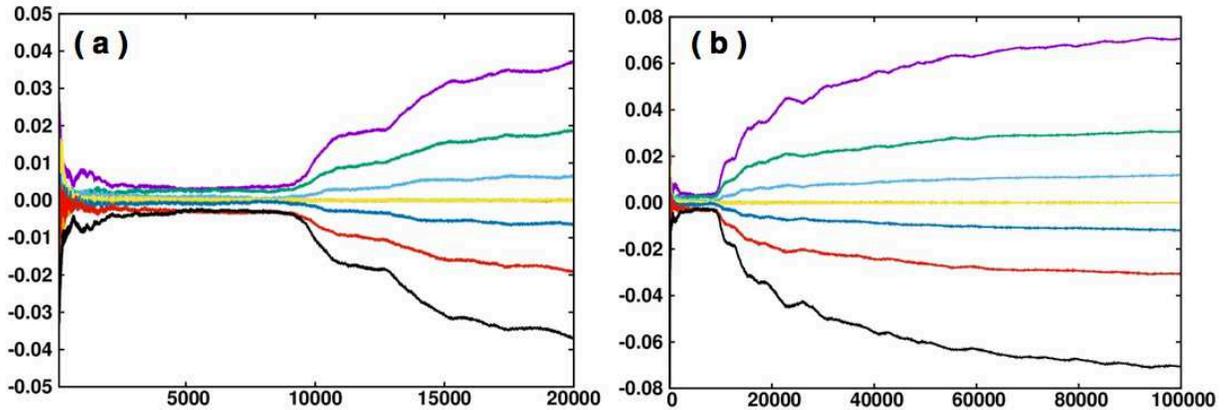}
\caption{
\noindent
Evolution ( a ) of the cumulative Lyapunov spectrum for the $\phi^4$ model from the initial condition
$\{ \ p \ \} = \{ \ 2,  2,  2,  -2 \ \}$ with $dt = 0.001$ and the offset length $\delta = 0.000001$ .
Simulations using RK4 or RK5 in either phase space or in tangent space all ``look'' similar in this
time range.
Evolution ( b ) of the Lyapunov spectrum over a longer time range showing pairing using RK4.
Results were calculated using Karl Travis' FORTRAN 90 computer program in the $\phi^4$ phase space.
}
\end{figure}

The simplest Lyapunov algorithm for the $\phi^4$ model is exactly that used with the springy pendula.  We
follow $D+1$ trajectories in the $D$-dimensional phase space, rescaling them at every timestep
to obtain the complete spectrum of $D = 8$ instantaneous Lyapunov exponents. This phase-space integration
of nine trajectories, followed by Gram-Schmidt orthonormalization, can be modified
by using Lagrange multipliers to impose the eight constant-length constraints and the $(1/2)(8\cdot7) =
28$ orthogonality constraints. A third approach, particularly simple to implement for the $\phi^4$
model with its power-law equations of motion, is to linearize the motion equations so that the
offset vectors, rather than being small, can be taken as unit vectors in ``tangent space''.  By
using separate integrators for the ``reference trajectory'' and for the eight unit vectors the
programming is at about the same level of difficulty as is that of the straightforward phase-space 
approach.  We implemented both approaches for the $\phi^4$ problems and found good agreement for
the Lyapunov spectra at a visual level, even for calculations using a billion timesteps.  This is
because the reference trajectories for the phase-space and tangent-space algorithms are identical.

\section{Useful Integration Techniques}

\noindent

\begin{figure}
\includegraphics[width=5.0in,angle=+90.]{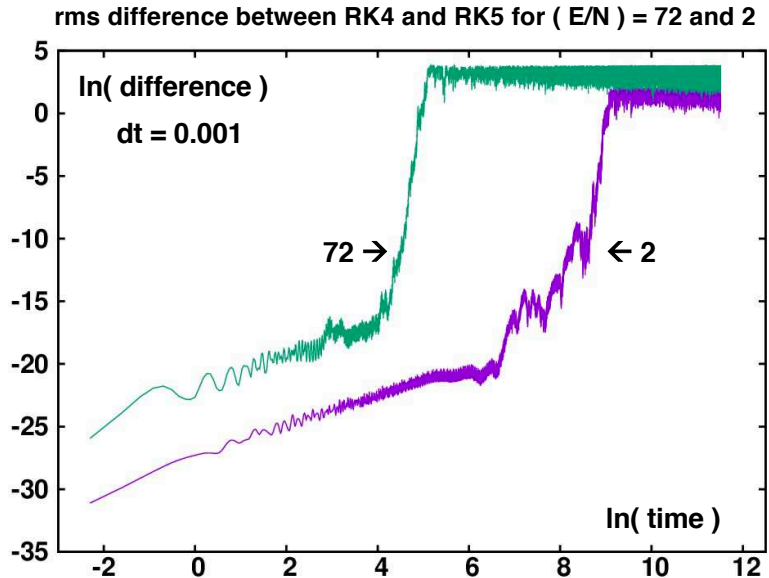}
\caption{
\noindent
We illustrate the dependence of integration errors ( RK4 {\it versus} RK5 )
for low energy and higher-energy $\phi^4$ chains with initial values $\{ \ q \ \} = \{ \ 0,0,0,0 \ \}$ with
$\{ \ p \ \} = \{ \ 2,2,2,-2 \ \} \ {\rm and} \ \{ \ 12,12,12,-12 \ \}$ respectively. After a linear
induction period the errors grow exponentially, reflecting Lyapunov instability, until saturation.  
}
\end{figure}

Fourth-order and fifth-order Runge-Kutta integrators are particularly useful algorithms for small
systems.  First, these integrators  are easy to program. These integrators are also explicit, a real
simplification whenever a variable timestep is desirable. Their errors are typically opposite in sign.
For the simple harmonic oscillator the fourth-order energy decays while the fifth-order energy diverges.
By choosing a sufficiently small timestep, for which the two algorithms agree, one can be confident in
the accuracy of the trajectories.  Another useful technique is adapative integration : comparing solutions
with a single timestep $dt$ to those from two successive half steps with $(dt/2)$.  The timestep is then
adjusted up or down
by a factor of two whenever it is necessary to keep the root-mean-squared error in a prescribed band,
$10^{-12} > {\tt error} > 10^{-14}$ for instance.\cite{b12}

At the expense of about a factor of fifty in computer time, FORTRAN makes it possible to carry out
quadruple-precision simulations with double-precision programming by changing the gnu compiler command :
$$
{\tt gfortran} \ {\rm -}{\tt O} \ {\rm -} {\tt o} \ {\tt xcode} \ {\tt code.f} \longrightarrow
{\tt gfortran} \ {\rm -}{\tt O} \ {\rm -} {\tt o} \ {\tt xcode} \
{\tt -}{\tt freal}{\rm -}{\tt 8}{\rm -}{\tt real}{\rm -}{\tt 16 \ code.f}
$$
Here the FORTRAN program is {\tt code.f} and the executable is {\tt xcode} .

\section{The 2017 Ian Snook Prize Problem}

The springy pendula and $\phi^4$ problems detailed here show that ``pairing'' is typically present after
sufficient time, with that time sensitive to the largest Lyapunov exponent as well as to the initial
conditions. There are several features of these introductory problems that merit investigation :\\
\noindent
[ 1 ] To what extent is there an unique chaotic sea ? Can the symmetry of the initial conditions limit the portion
of phase space visited when the dynamics is chaotic ?\\
\noindent
[ 2 ] Within the $\phi^4$ model's chaotic sea do
the time-averaged kinetic temperatures  $\{ \ T_i = \langle \ p_i^2 \ \rangle \ \}$ , agree for all the particles ?
( If not, a thermal cycle applying heat and extracting work from the chain could be developed so as to
violate the Second Law.\cite{b11} )\\
\noindent
[ 3 ] Is the pairing time simply related to the Lyapunov exponents and the chain length ?\\
\noindent
[ 4 ] Is the accuracy of the pairing simply related to the accuracy of the integrator ?\\
\noindent
The next and last question, which motivated this year's Prize Problem seems just a bit more difficult : [ 5 ]
Can relatively-simple autonomous Hamiltonian systems be devised for which long-time local pairing is
absent ? Our exploratory work has suggested that dynamical disturbances induced by collisions, with those
collisions separated by free flight, could lead to repeated violations of pairing\cite{b13,b14}. On the other
hand Dettmann and Morriss have published a proof of pairing for isokinetic systems\cite{b15}.  A simple gas of
several diatomic or triatomic molecules is likely to be enough to settle that question.

The 2017 Ian Snook Prize will be awarded to the most interesting paper discussing and elucidating these
questions.  Entries should be submitted to Computational Methods in Science and Technology, cmst.eu, prior
to 1 January 2018. The Prize Award of 500 United States dollars sponsored by ourselves, and the Additional
Ian Snook Prize Award, also 500, will be awarded to the author(s) of the paper best addressing
this Prize Problem.

\section{Acknowledgments}

We are grateful to the Poznan Supercomputing and Networking Center for their support of these prizes
honoring our late Australian colleague Ian Snook (1945-2013).  We also appreciate useful comments,
suggestions, and very helpful numerical checks of our work furnished by Ken Aoki, Carl Dettmann, Clint
Sprott, Karl Travis, and Krzysztof Wojciechowski.  We particularly recommend Aoki's reference 10 for a
comprehensive study of the dynamics of one-dimensional equilibrium $\phi^4$ systems.

\end{document}